# Uncertainty-Cognizant Model Predictive Control for Energy Management of Residential Buildings with PVT and Thermal Energy Storage

Hossein Kalantar-Neyestanaki, Madjid Soltani

*Abstract*—**The building sector accounts for almost 40 percent of the global energy consumption. This reveals a great opportunity to exploit renewable energy resources in buildings to achieve the climate target. In this context, this paper offers a building's energy system embracing a heat pump, a thermal energy storage system along with grid-connected photovoltaic thermal (PVT) collectors to supply both electric and thermal energy demands of the building with minimum operating cost. To this end, the paper develops a stochastic model predictive control (MPC) strategy to optimally determine the set-point of the whole building's energy system while accounting for the uncertainties associated with the PVT energy generation. This system enables the building to 1-shift its electric demand from high-peak to off-peak hours and 2- sell electricity to the grid to make energy arbitrage.**

## I. INTRODUCTION

**B**uilding sector plays an important role in the global warming and clime change. This sector is responsible for around 40% of the global energy consumption, and accordingly, 40% of energy-related global CO2 emissions [1,2,3]. To meet climate goals, it is of paramount importance to modernize building energy management systems (BEMSs) and integrate renewable energy resources (RES) in this sector with the purpose of supplying both thermal and electric demands of buildings [4]. In this respect, new policies like the Directive on Energy Performance of Buildings have set themselves the goal of reaching fully decarbonized building sector by 2050 [5,6]. However, these new policies cannot be realized without novel BEMSs. For example, all promoting incentives and policies for increasing the penetration rate of RES in building sector have not been significantly effective and currently less than 15% of the total energy demand of buildings is provided by RES installed in them [7]. To quantify the compliance of the buildings with the fully decarbonized target, an index called building's self-consumption [8,9], or more specifically self-consumption of RES in the building, can be defined with value between 0 and 1. This index indicates the proportion of the building's total energy demand (including both thermal and electric demand) over a selected time horizon that is supplied by the RES installed in the building. The main obstacles in the way of reaching fully decarbonized buildings (i.e. buildings with self-consumption equal to 1) are that:

- RES's generated energy is fully dependent on the meteorological conditions and does not follow the building's energy demand. More specifically, there is a time mismatch between supply and demand of energy. Above all, RES's generated energy might be accompanied by spikes.
- The demand of buildings is volatile and accompanied with spikes (of demand). These spikes of demand might cause congestion in the electric network during peak hours. Therefore, shifting demand (i.e. moving the consumption from peak to off-peak hours) and removing spikes of demand are vital to keep the security of the electricity supply.

To deal with this issue, the capability of thermal energy storage systems (TESSs) for storing energy can be leveraged to 1-store energy when there is a surplus of RES's energy generation and 2- deliver energy when there is a shortage of RES's energy generation. On the other hand, the TESSs can help to accomplish demand-shifting, i.e. shifting energy consumption from peak to off-peak hours in buildings [10], [11].

The existing literature have broadly used a building energy system architecture consisting of a TESS along with RES to improve the self-consumption of buildings, thereby, facilitate the transition of the building sector towards a decarbonized sector. In this emerging field, the fundamental question that naturally arises is that:

*How should be the building's energy system managed to facilitate the integration of RES in the building with the purpose of supplying both thermal and electric demands of the building while minimizing the building's energy cost and preserving the thermal comfort of the residents?*

The ASHRAE handbook [12,13] provides a comprehensive survey about the control techniques used in BEMSs. It highlights that classical control techniques have been widely used in BEMS owing to their simplicity in design and low computational complexity. The heating, ventilation and air conditioning (HVAC) systems are traditionally controlled using a classical control method called Rule-Based Controllers

Hossein Kalantar-Neyestanaki is with Niroo Research Institute (NRI), Tehran, Iran (hosein.kalantar@gmail.com).

Madjid Soltani is with Mechanical Engineering Department of K.N.Toosi University of Technology, Tehran, Iran (msoltani@kntu.ac.ir).

(RBC). RBC divides the whole building into several parts and accordingly manages each part separately. The simplest RBC is On/Off switches installed in each room of the old buildings to control the heating and cooling system of the room. To achieve a more efficient heating and cooling system, modern buildings rely on Proportional-Integral-Derivative (PID) control loops to control variable frequency drives. Although highly sophisticated classical controllers might provide a good performance in room-level applications, they are never able to support building-level applications like optimizing the whole energy consumption of the building [14].

Afram et al. [15] classified HVAC control techniques into several main categories: 1-classical control, 2- soft control, 3-hard control, and 4-other minor control techniques. Among all mentioned control techniques, model predictive control (MPC), belonging to the hard control category, is a well-established method that has been greatly receiving researchers' attention. In contrast with the classical control techniques, MPC provides a holistic perception about the whole building over a selected time horizon. More specifically, MPC exploits the existing powerful prediction approaches to predict unknown parameters (such as outdoor temperature, energy demand of the building, energy generation of the RES installed in the building) of the system over the selected time horizon. Then, MPC determines the optimal control actions by solving an optimization problem whose constrains model technical restrictions of the building's energy system over the whole time horizon [16]. On the other hand, the existing agile commercial optimization solvers like Gurobi and CPLEX along with advanced monitoring systems installed in the modern buildings make the MPC as a practical approach to optimally steer the energy system of the modern buildings embracing RES, TESS and HVAC [14]. The existing literature have mainly employed MPC to either 1- optimally control a single HVAC system (e.g. TESS, ground-source heat pump, window control, etc.) or 2- implement demand response (DR) in the buildings. DR refers to managing the building energy consumption in order to alleviate congestion in the electric power grid, such as reducing or shifting consumption at peak hours.

Halvgaard et al. [17] applied an MPC scheme to optimally operate an electric heating element coupled with a TESS while covering the mismatch between the generated thermal energy of the solar collector and the thermal demand of the building. This work experimentally represented that its developed MPC strategy can achieve 25-30 % annual energy cost saving in comparison with the thermostat control strategy. Deng et al. [18] constructed an MPC strategy to derive the optimal set-points of a central chiller plant paired with TESS. It relied on a heuristic algorithm to solve the MPC problem. Its presented results revealed that the MPC scheme is able to significantly decrease the electricity cost, i.e. 10.84% electricity cost saving. D'Ettorre et al. [19] investigated the potential cost saving that can be achieved by installing a TESS in a hybrid thermal system composed of a heat pump and a gas boiler. This work adopted an MPC scheme to find the optimum size of the TESS and optimally control the hybrid system. Its obtained results illustrated energy cost saving up to 8% compared to the case without TESS. Carvalho et al. [20] considered a ground source heat pump (GSHP) to provide the thermal demand of a building while minimizing the electricity cost. To this end, it leveraged the thermal storage capacity of the building and offered a load shifting strategy where the GSHP consumes electricity in off-peak hours and pre-heat the building. In this way, it shifts the electric demand of the building and accordingly minimizes the building's electricity cost. Finally, its presented results showed 34% reduction in the building's electricity cost. Arteconi et al. [10] considered a building energy system comprised of a TESS and a heat pump with either radiators or underfloor heating system. It showed that the stored energy of the TESS can be utilized to preserve the building's thermal comfort even if the heat pump is turned off for 3 hours during peak hours. This work taped this unique capability of TESS (i.e. storing energy) to flatten the shape of the building's electricity demand curve by switching off the heat pump during peak hours, thereby, reducing electricity cost considering the time-of-use (TOU) tariff. Baniasadi et al. [21] considered a building equipped with two TESSs and a water storage tank (WST). To empower this building to provide DR service to the electric grid, this work firstly defined a dynamic temperature set-point on the basis of the real-time electricity price, thereby, enhancing peak-load shifting of demand. Then, it developed an MPC to optimally determine the thermal energy that the GSHP should provide to the WST to minimize the operation cost of the entire energy system and maintain the indoor temperature within a desirable comfort range. Touretzky et al. [22] proposed an economic MPC in the context of buildings equipped with a TESS, in which the building HVAC system was controlled with the aim of minimizing the purchased electricity cost while considering TOU pricing. Shah et al. [23] presented an approach to investigate the effectiveness of using TESS in reducing the cost of providing hot water for a residential building for 24 hours while considering TOU pricing.

In sum, there are numerous researches related to the application of MPC in building. All these works highlight the importance of accurate modeling of the building energy system, above all they bear testimony to the capability of MPC for achieving significant amount of financial/energy saving without compromising the comfort level of residents [24,25,26]. Although a variety of approaches have been developed to manage the building's energy system, the existing literature however lacks a framework to treat the question put forward earlier. In this context, the main contributions of the paper can be enumerated as follows:

- Offering linear models for all elements of the building energy system, most notably the building and TESS. These linear models are leveraged to offer a computationally affordable MPC strategy.

- Exploiting a PVT-air dual source heat pump to recover energy from both solar irradiation and ambient air, thereby improving the performance of the heat pump under a wide range of operating conditions.

- Introducing a machine learning-based approach to exploit the existing historical data, thereby, predicting the RES generation (solar irradiation) and outdoor ambient air temperature over the selected tine horizon.

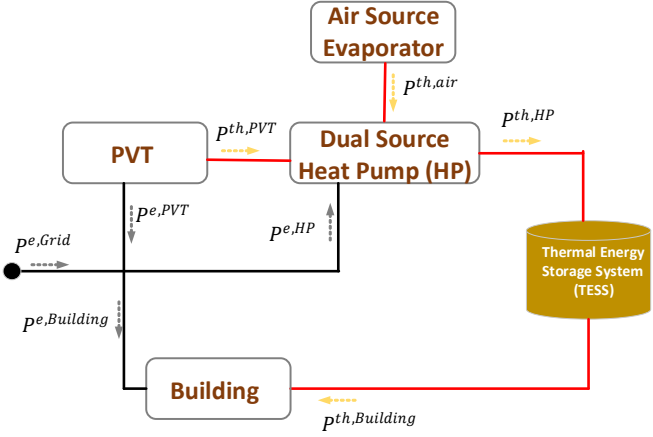

Fig. 1 Energy flow between the main components of the Building's energy system.

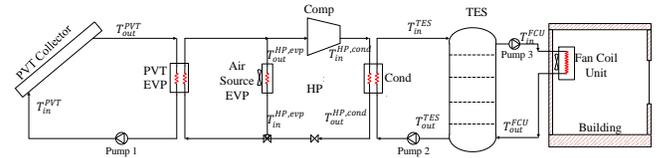

Fig. 2 Detailed model of the building's energy system.

Moreover, this approach is able to model the uncertainties associated with RES generation (solar irradiation) through a number of scenarios.

- Constructing a stochastic MPC strategy to maximize the building's self-consumption while accounting for the uncertainties associated with RES. It is worth highlighting that this MPC strategy takes into account the temporal variations of the electricity price (i.e. TOU tariff), RES generation, outdoor ambient air temperature over the selected time horizon.

The rest of the paper is organized as follows. Section II firstly describes the architecture of the building 's energy system and then models all elements of the building's energy system. section III develops a stochastic MPC strategy to optimally steer the building's energy system while accounting for time-variant electricity price, outdoor temperature, solar irradiation as well as uncertainties associated with solar irradiation. Section IV introduces the case study and finally section V presents simulation results followed by the conclusions.

## II. SYSTEM DESCRIPTION AND MODELLING

### A. System Architecture

This paper considers an energy system formed of 3 main components, namely a thermal energy storage system (TESS), a PVT-air dual source heat pump and a photovoltaic thermal (PVT) collector, to supply both thermal and electric demands of a building. The schematic block diagram of this energy system is shown in Fig. 1. As it can be seen, the absorbed electricity from the grid ($P^{e,Grid}$) along with the generated electricity of the PVT collector ($P^{e,PVT}$) are leveraged to satisfy both the electric demand of the building ($P^{e,Building}$) and the electric power consumption of the heat pump ($P^{e,HP}$). More specifically, the heat pump consumes electric power ($P^{e,HP}$) to tackle two fundamental tasks:

- The absorbed thermal energy from the PVT evaporator ($P^{th,PVT}$) doesn't have an appropriate temperature level to be directly delivered to the TESS. To overcome this issue, the heat pump receives the thermal energy of the evaporators and consumes electric power ($P^{e,HP}$), thereby providing thermal energy ($P^{th,HP}$) with temperature level that can be delivered to the TESS.
- During periods (like nights) where the PVT cannot generate any thermal energy, the heat pump uses air source evaporator

and consumes electric power ($P^{e,HP}$) to supply the required thermal demand of the TESS i.e. $P^{th,HP}$.

The TESS has a paramount role to constantly keep a balance between the supply and demand of the thermal energy in the building, thereby preserving the thermal comfort of the residents. More specifically, the TESS discharges (and respectively charges) its stored thermal energy when the thermal demand of the building, i.e. $P^{th,Building}$ is larger (and respectively smaller) than the thermal power delivered by the heat pump, i.e. $P^{th,HP}$.

It is worth highlighting that this paper takes into account the temporal variations of the electricity price over the selected time horizon. Therefore, the best strategy to significantly decrease the operation cost of the building's energy system is to shift the electricity consumption of the system ($P^{e,Grid}$) from periods with high prices to periods with low prices, i.e. from peak to off-peak periods. This strategy can be achieved thanks to the TESS:

- **During periods with low prices**: the heat pump stores thermal energy in the TESS, i.e. the thermal energy that heat pump delivers to the TESS ($P^{th,HP}$) is larger than the thermal demand of the building ($P^{th,Building}$).
- **During periods with high prices**: the TESS discharges its stored thermal energy; therefore, the heat pump can consume less electricity and even stop working.

In other words, the TESS allows the heat pump to mainly operate in periods where the price of the electricity is low. Therefore, the TESS helps to minimize the operation cost of the system, i.e. the cost associated with purchasing electricity from the grid ($P^{e,Grid}$), over the whole defined time horizon. The detailed model of the building's energy system is illustrated in Fig. 2Fig. 2. The following sub-sections elaborate on the modeling of each part of the building's energy system.

### B. The Building's Thermal Model

The BEMS, presented in this paper, leverages the building's thermal model to determine the evolution of the building's indoor air temperature over the time, thereby, ensuring the thermal comfort of the residents while optimally steering the building's energy system. In this respect, this sub-section sets out to derive the thermal model of a typical building in time domain. The main thermal interactions affecting on the building's indoor air temperature, shown in Fig. 3, are:

- heat input from solar radiation;
- heating or cooling loads from fan coil;
- heat transfer through the walls and windows;
- internal heat gain corresponding to the activities of occupants such as metabolic heat, utilization of electrical devices, and thermal emission of artificial lighting.

This paper considers all elements located inside the building as a part of the building's internal thermal mass which is shown as a part of the floor in Fig. 3. Therefore, the building's indoor space can be considered as two different thermal zones, i.e. the

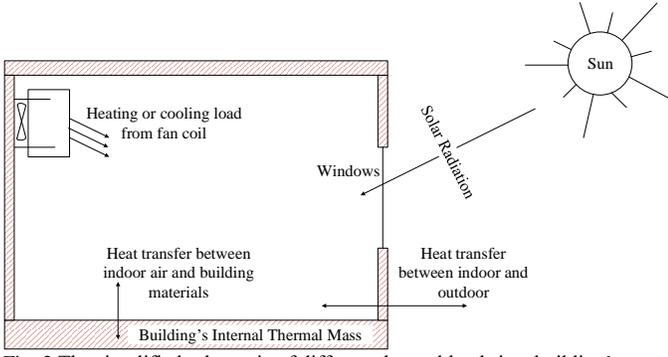

Fig. 3 The simplified schematic of different thermal loads in a building's zone.

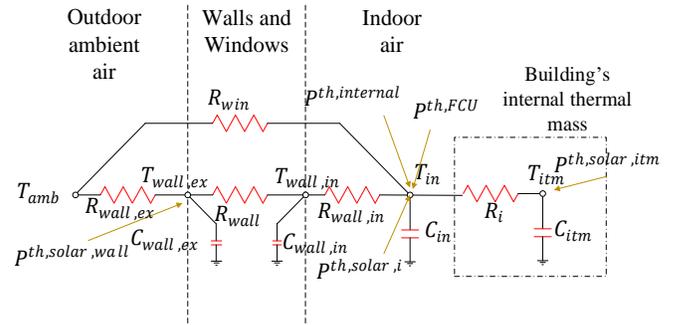

Fig. 4 Schematic of a building's thermal network.

air and the building's internal thermal mass. Then, the equivalent thermal-electrical model of the building is extracted as shown in Fig. 4. The privileged features of this model are:

1- this model simply reflects the key thermal characteristics of the building without increasing the complexity of the model, thereby, having low computational burden;

2- this model can predict the system behavior under different conditions;

3- this model is applicable for both short-term and long-term studies where the variations of the building's indoor air temperature under dynamic situations are investigated.

The main components of the building's thermal network and heat fluxes are shown in Fig. 4. This thermal network consists of 4 parts that are 1-outdoor environment, 2-building envelope (i.e. walls and windows), 3-indoor air, and 4-building's internal thermal mass. In this model, each node is described by its temperature (T) and its thermal capacity (C). Additionally, the heat transfer between two nodes are modeled by thermal resistances (R) [27].

The heat transfer on the external wall arises from the convective heat transfer with the outdoor ambient air, not to mention radiative heat transfer with the sky ($P^{th,solar,wall}$). The solar radiation enters the building through the windows and its accompanied energy is transmitted to the indoor air ($P^{th,solar,i}$) and the building's internal thermal mass ($P^{th,solar,itm}$). The building's internal thermal mass is modeled by a thermal resistance ($R_i$) and a thermal capacitance ($C_{itm}$). The building's indoor air absorbs:

1- some portion of the solar radiation entered the building through the windows ($P^{th,solar,i}$);

2- thermal loads of the equipment and occupants ($P^{th,internal}$);

3- thermal load of the fan coil ($P^{th,FCU}$).

The abovementioned thermal loads are transmitted to the building's internal thermal mass through thermal resistance $R_i$ and accordingly stored in it. This stored thermal energy is gradually released to the building's indoor air in the next time intervals. Relying on the presented thermal-electrical model, the thermal dynamics of the building can be mathematically modeled as follows:

The energy balance for the exterior wall surface:

$$C_{wall,ex} \frac{dT_{wall,ex}}{dt} = \frac{T_{amb} - T_{wall,ex}}{R_{wall,ex}} - \frac{T_{wall,ex} - T_{wall,in}}{R_{wall}} + P^{th,solar,wall} \quad (1)$$

The energy balance for the interior wall surface:

$$C_{wall,in} \frac{dT_{wall,in}}{dt} = \frac{T_{wall,ex} - T_{wall,in}}{R_{wall}} - \frac{T_{wall,in} - T_{in}}{R_{wall,in}} \quad (2)$$

The energy balance for the indoor air:

$$C_{in} \frac{dT_{in}}{dt} = \frac{T_{amb} - T_{in}}{R_{win}} + \frac{T_{wall,in} - T_{in}}{R_{wall,in}} + \frac{T_{itm} - T_{in}}{R_i} + P^{th,internal} + P^{th,FCU} + P^{th,solar,i} \quad (3)$$

The energy balance for the internal thermal mass:

$$C_{itm} \frac{dT_{itm}}{dt} = \frac{T_{in} - T_{itm}}{R_i} + P^{th,solar,itm} \quad (4)$$

In expressions (1) to (4), parameters C and R respectively represent the thermal capacitance and thermal resistance of the respective parts, which depend on the building construction. $P^{th,solar,wall}$ indicates the solar thermal power absorbed by the external wall surfaces. It can be calculated as:

$$P^{th,solar,wall} = \alpha . G . A_{wall} \quad (5)$$

where $\alpha$ is the absorptance coefficient of the wall's surface, $G$ is the solar irradiance ($\frac{W}{m^2}$), and $A_{wall}$ is the area of the external wall in $m^2$. $P^{th,solar}$ indicates the solar thermal power entered the building through the windows and can be calculated as:

$$P^{th,solar} = A_{e,wind} . G . SHGC \quad (6)$$

where $A_{e,wind}$ is the effective windows area in $m^2$, $G$ is the solar irradiance ($\frac{W}{m^2}$), and solar heat gain coefficient (SHGC) is a parameter indicating the fraction of the solar thermal power entered the building through windows, either transmitted directly or absorbed and subsequently released as heat inside the building. $P^{th,solar}$ can be divided into two terms $P^{th,solar,i}$ and $P^{th,solar,itm}$ which represent the solar thermal power absorbed by the building's indoor air and internal thermal mass, respectively. The former term, i.e. $P^{th,solar,i}$, affects on the indoor air temperature while the latter term, i.e. $P^{th,solar,itm}$, affects on the internal mass temperature. Relying on expression (6), $P^{th,solar,i}$ and $P^{th,solar,itm}$ can be easily calculated as:

$$P^{th,solar,i} = f . A_{e,wind} . G . SHGC \quad (7)$$
$$P^{th,solar,itm} = (1 - f) . A_{e,wind} . G . SHGC \quad (8)$$

where $f$ and $(1 - f)$ indicate the fraction of $P^{th,solar}$ that is transmitted to the indoor air and to the internal thermal mass, respectively.

$P^{th,FCU}$ is the total cooling or heating thermal power supplied by the fan coil and $P^{th,internal}$ is the internal heat gain (associated with occupants, lights and equipment).

Expressions (1)-(8) all together define a set of first order differential equations governing the evolution of the building's temperatures ($T_{wall,ex}, T_{wall,in}, T_{itm}, T_{in}$) over the time. This set of equations can be rearranged to achieve the state-space model describing the dynamics of the building's temperatures as:

$$\frac{dX}{dt} = a . X + b . u + e . d \quad (9)$$

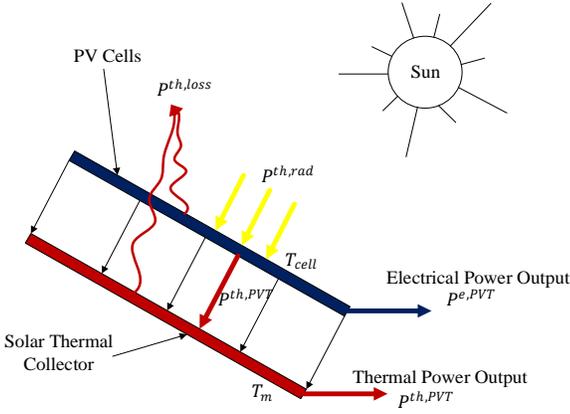

Fig. 5 The simplified diagram of a PVT collector and its corresponding generated thermal and electrical powers.

where $X = [T_{wall,ex}, T_{wall,in}, T_{itm}, T_{in}]^T$ is the system state vector, $u = [P^{th,FCU}]$ is the control input vector, $d = [G, T_{amb}, P^{th,internal}]^T$ is the disturbance vector. $a$ , $b$ and $e$ respectively represent the system matrix, the input matrix and the disturbance matrix. They can be calculated as ((10) shown at the bottom of the page):

$$b = \begin{bmatrix} 0 \\ 0 \\ 0 \\ \frac{1}{C_{in}} \end{bmatrix} \tag{11}$$

$$e = \begin{bmatrix} \frac{\alpha \cdot A_{wall}}{C_{wall,ex}} & \frac{1}{R_{wall,ex} \cdot C_{wall,ex}} & 0 \\ 0 & 0 & 0 \\ \frac{(1-f) \cdot A_{e,wind} \cdot SHGC}{C_{itm}} & 0 & 0 \\ \frac{f \cdot A_{e,wind} \cdot SHGC}{C_{in}} & \frac{1}{R_{win} \cdot C_{in}} & \frac{1}{C_{in}} \end{bmatrix} \tag{12}$$

In order to embed the thermal model of the building in the MPC problem, the continuous-time state space model, presented in (9)-(12), is discretized and converted to the discrete-time state-space model:

$$X_{k+1} = a_d \cdot X_k + b_d \cdot u_k + e_d \cdot d_k \tag{13}$$

where $a_d$, $b_d$, and $e_d$ are the corresponding matrixes of the discrete-time state space model of the building [28,29,30].

## C. The PVT Collector's Thermal and Electrical Models

Photovoltaic-thermal (PVT) collectors are able to simultaneously convert solar energy (accompanied with solar irradiation) into electrical and thermal energy as schematically represented in Fig. 5. In this context, this sub-section first presents the thermal and then the electrical model of the PVT collector.

### Thermal model of the PVT collector

The equivalent thermal-electrical model of the PVT collector can be derived by considering 1-PV cell and 2-solar thermal collector as nodes of the model. As shown in Fig. 6, the PVT collector is modeled as a two-node thermal network whose two nodes are connected together via the internal thermal resistance of the PVT collector ($R_{PVT}$). An effective thermal heat capacity ($C_{eff}$) connected to the solar thermal collector node ($T_m$) is considered to model thermal capacities of cooling fluid, absorber, PVT structure and frame, and insulations all together [31]. It should be noted that each of these two nodes are described by their temperature. Therefore, PV cell node (and respectively solar thermal collector node) is represented by its temperature, i.e. $T_{cell}$ ($T_m$).

The energy balance for the entire PVT collector results in:

$$c_{eff} A_{PVT} \frac{dT_m}{dt} = P^{th,rad} - P^{th,loss} - P^{th,PVT} \tag{14}$$

where $T_m$ is the mean temperature of the cooling fluid:

$$T_m = \frac{T_{in}^{PVT} + T_{out}^{PVT}}{2} \tag{15}$$

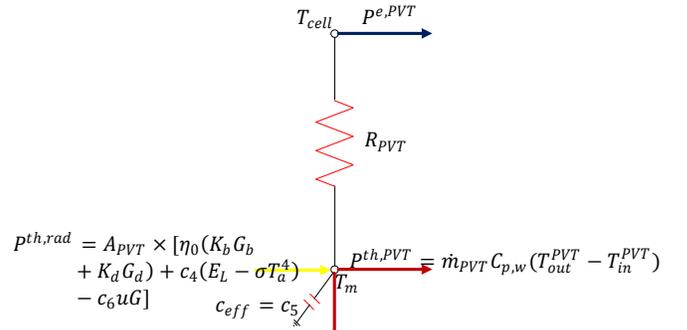

$P^{th,rad} = A_{PVT} \times [\eta_0(K_b G_b + K_d G_d) + c_4(E_L - \sigma T_a^4) - c_6 u G]$

$c_{eff} = c_5$

$P^{th,loss} = A_{PVT} \times [c_1.(T_m - T_a) + c_2.(T_m - T_a)^2 + c_3.u(T_m - T_a)]$

Fig. 6 Thermal network of a PVT collector.

$$a = \begin{bmatrix} -\frac{1}{C_{wall,ex} \cdot R_{wall,ex}} - \frac{1}{C_{wall,ex} \cdot R_{wall}} & \frac{1}{R_{wall,ex} \cdot C_{wall}} & 0 & 0 \\ \frac{1}{R_{wall} \cdot C_{wall,in}} & -\frac{1}{R_{wall} \cdot C_{wall,in}} - \frac{1}{R_{wall,in} \cdot C_{wall,in}} & 0 & \frac{1}{R_{wall,in} \cdot C_{wall,in}} \\ 0 & 0 & -\frac{1}{R_i \cdot C_{itm}} & \frac{1}{R_i \cdot C_{itm}} \\ 0 & \frac{1}{R_{wall,in} \cdot C_{in}} & \frac{1}{R_{in} \cdot C_{in}} & -\frac{1}{R_{win} \cdot C_{in}} - \frac{1}{R_{wall,in} \cdot C_{in}} - \frac{1}{R_i \cdot C_{in}} \end{bmatrix} \tag{10}$$

and $P^{th,rad}$ represents the amount of the radiative thermal power that PVT collector receives, $P^{th,loss}$ represents the thermal power losses due to conduction and convection heat transfer with the ambient air, and $P^{th,PVT}$ represents the generated thermal power of the PVT collector that is conveyed to the cooling fluid, i.e. circulating water. $P^{th,PVT}$, $P^{th,rad}$ and $P^{th,loss}$ are mathematically characterized in the following.

$$P^{th,PVT} = \dot{m}_{PVT}.C_{p,w}.(T_{out}^{PVT} - T_{in}^{PVT}) \qquad (16)$$

where $\dot{m}_{PVT}$ is cooling fluid mass flow rate, $C_{p,w}$ specific thermal capacity, and $T_{out}^{PVT}$ and $T_{in}^{PVT}$ are temperature of outlet and inlet fluid from the PVT collector, respectively.

$$P^{th,rad} = A_{PVT}\eta_0(K_bG_b + K_dG_d) + A_{PVT}c_4(E_L - \sigma T_{amb}^4) - A_{PVT}c_6uG \qquad (17)$$

where $A_{PVT}$ is the surface area of the collector, $\eta_0$ the zero-loss collector efficiency, $K_b$ the incidence angle modifier (IAM) for solar beam radiation $G_b$, $K_d$ the IAM for diffuse radiation $G_d$, $c_4$ the sky temperature dependency on the long wave radiation exchange, $E_L$ the long wave irradiance, $\sigma$ the Stefan-Boltzmann constant, $c_6$ the wind speed dependency of the zero loss efficiency, $u$ wind speed over the surface of the collector and $G$ the global irradiance over the collector surface [31]. $K_b$ is calculated as:

$$K_b = 1 - b_{0,th}.\left[\left(1/_{cos\theta}\right) - 1\right] \qquad (18)$$

where $b_{0,th}$ is the constant for the thermal incidence angle modifier and $\theta$ the incidence angle of the beam radiation. $K_d$ is similarly determined on the basis of (18) except that in this case the diffuse radiation must be considered instead of solar beam radiation.

$$P^{th,loss} = A_{PVT}c_1(T_m - T_{amb}) + A_{PVT}c_2(T_m - T_{amb})^2 + A_{PVT}c_3u(T_m - T_{amb}) \qquad (19)$$

where $T_{amb}$ is the outdoor ambient temperature, $c_1$ the heat loss coefficient, $c_2$ the temperature dependency of the heat loss coefficient and $c_3$ the wind speed dependency of the heat loss coefficient, $u$ wind speed over the surface of the collector [31].

Substituting $T_m$, $P^{th,PVT}$, $P^{th,rad}$ and $P^{th,loss}$ from (15), (16), (17) and (19) into (14) results in the following expression that characterized $P^{th,PVT}$ [32]:

$$\frac{A_{PVT}\,c_{eff}}{2}\left(\frac{dT_{in}^{PVT}}{dt} + \frac{dT_{out}^{PVT}}{dt}\right) =$$
$$A_{PVT}\eta_0(K_bG_b + K_dG_d)$$
$$-A_{PVT}c_1\left(\frac{T_{in}^{PVT} + T_{out}^{PVT}}{2} - T_{amb}\right)$$
$$-A_{PVT}c_2\left(\frac{T_{in}^{PVT} + T_{out}^{PVT}}{2} - T_{amb}\right)^2$$
$$-A_{PVT}c_3u\left(\frac{T_{in}^{PVT} + T_{out}^{PVT}}{2} - T_{amb}\right)$$
$$+A_{PVT}c_4(E_L - \sigma T_{amb}^4)$$
$$-A_{PVT}c_6uG$$
$$-\dot{m}_{PVT}.C_{p,w}.(T_{out}^{PVT} - T_{in}^{PVT}) \qquad (20)$$

The energy balance for the node corresponding to the PV cell ($T_{cell}$) results in:

$$T_{cell} = T_m + R_{PVT}\frac{P^{th,PVT}}{A_{PVT}} \qquad (21)$$

where $T_{cell}$ indicates the temperature of PV cell ($T_{cell}$), $R_{PVT}$ is the internal thermal resistance of the PVT collector, and $P^{th,PVT}$ output thermal power of the PVT collector.

Substituting $T_m$ and $P^{th,PVT}$ from (15) and (16) into (21) yields:

$$T_{cell} = \frac{T_{in}^{PVT} + T_{out}^{PVT}}{2} + \frac{R_{PVT}.\dot{m}_{PVT}.C_{p,w}}{A_{PVT}}(T_{out}^{PVT} - T_{in}^{PVT}) \qquad (22)$$

### Electrical model of the PVT collector

The operating condition of the PVT collector might differ from the reference (nominal) operating condition defined in the datasheet of the PVT collector. Accordingly, the electrical efficiency of the PVT collector might vary from its nominal efficiency, i.e. $\eta_{el,ref}$ that is stated in the datasheet of the PVT collector. To take into account the impact of the operating condition on the electrical efficiency of the PVT collector, its efficiency can be modified as:

$$\eta_{el} = \eta_{el,ref}.PR_{total} \qquad (23)$$

where $\eta_{el,ref}$ indicates the electrical efficiency at the reference condition (nominal efficiency) stated in the datasheet of the PVT collector. $PR_{total}$ is the overall instantaneous performance ratio and can be calculated as:

$$PR_{total} = PR_{IAM}.PR_G.PR_T \qquad (24)$$

where $PR_{IAM}$, $PR_G$ and $PR_T$ are:

#### $PR_{IAM}$: Loss effects of incidence angle

The electrical efficiency of the PVT collector is affected by the incident angle of the beam radiation. This effect is quantified as [33]:

$$PR_{IAM} = 1 - b_{0,el}.\left[\frac{1}{cos\theta} - 1\right] \qquad (25)$$

where $b_{0,el}$ is the constant for electrical incidence angle modification (IAM). The value of $b_{0,el}$ can be determined either by measurement or set equal to the thermal IAM i.e. $b_{0,th}$. $\theta$ is the incidence angle of beam radiation.

#### $PR_G$: Loss effects of irradiance

The performance ratio due to irradiance losses ($PR_G$) is quantified as [34]:

$$PR_G = a.G + b.\ln(G + 1) + c.\left[\frac{(\ln(G + e))^2}{G + 1} - 1\right] \qquad (26)$$

where a, b, and c are model parameters, G is solar irradiance and e is Euler's number.

#### $PR_T$: Loss effects of PV cell temperature

The electrical efficiency of the PVT collector is significantly affected by the PV cell temperature. This effect is quantified as [35]:

$$PR_T = 1 - \beta.(T_{cell} - T_{ref}) \qquad (27)$$

where $\beta$ is the power-temperature coefficient of the PV cell, $T_{cell}$ and $T_{ref}$ indicate the temperature of the PV cell in actual operating condition and reference operating condition, respectively. It should be noted that parameters $\eta_{el,ref}$ and $\beta$ can be extracted from the datasheet of the PVT collector.

Finally, the electrical power that the PVT collector generates ($P^{e,PVT}$) is calculated as:

$$P^{e,PVT} = \eta_{el,ref}.PR_{total}.G.A_{PVT} \qquad (28)$$

### D. The Thermal Energy Storage System's (TESS's) model

A well-known method for modeling of the TESS is the stratification approach [36,37,38,39] that vertically segments the TESS into several segments, i.e. layers. This paper relies on the stratification approach and considers three segments to

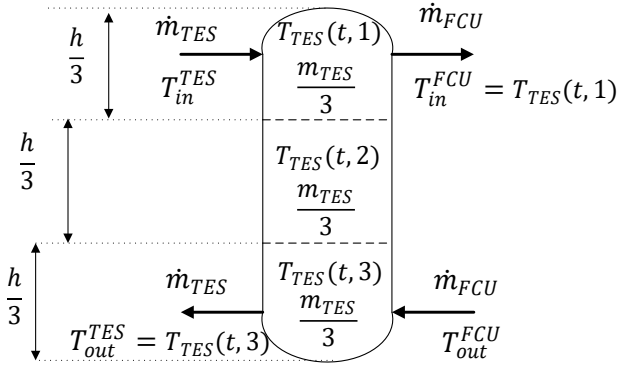

Fig. 7. The simplified schematic of TESS based on the stratification model.

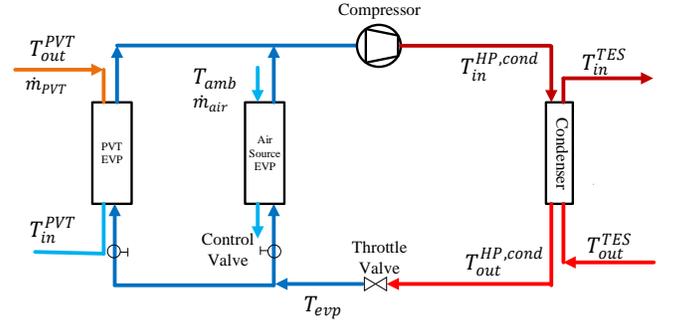

Fig. 8. The dual-source heat pump configuration.

model the TESS, as illustrated in Fig. 7. Each segment $i$ ($=1,2,3$) has a constant mass ($m_{TES}/3$) and a variable temperature ($T_{TES}(t,i)$). The internal thermal energy of each segment is affected by 1-the received thermal power from the heat pump, 2-the delivered thermal power to the building, 3-the heat losses through the surface area of the TESS tank as well as 4-the conduction heat transfer with the neighboring segments. The energy balance for three segments of the TESS can be expressed as:

First segment ($i = 1$):

$$\frac{m_{TES}}{3} \cdot c_{p,w} \cdot \frac{T_{TES}(t+1,1) - T_{TES}(t,1)}{\Delta t} =$$
$$\dot{m}_{TES} \cdot c_{p,w} \cdot [T_{in}^{TES}(t) - T_{TES}(t,1)]$$
$$-k_{TES} \cdot \frac{A_{TES}}{3} \cdot [T_{TES}(t,1) - T_{amb}(t)]$$
$$-k_w \cdot A_{CS} \cdot \left[\frac{T_{TES}(t,1) - T_{TES}(t,2)}{h/3}\right] \quad (29)$$

Second segment ($i = 2$):

$$\frac{m_{TES}}{3} \cdot c_{p,w} \cdot \frac{T_{TES}(t+1,2) - T_{TES}(t,2)}{\Delta t} =$$
$$(\dot{m}_{TES} - \dot{m}_{FCU}) \cdot c_{p,w} \cdot [T_{TES}(t,1) - T_{TES}(t,2)]$$
$$-k_{TES} \cdot \frac{A_{TES}}{3} \cdot [T_{TES}(t,2) - T_{amb}(t)]$$
$$+k_w \cdot A_{CS} \cdot \left[\frac{T_{TES}(t,1) - T_{TES}(t,2)}{h/3}\right.$$
$$\left. - \frac{T_{TES}(t,2) - T_{TES}(t,3)}{h/3}\right] \quad (30)$$

Third segment ($i = 3$):

$$\frac{m_{TES}}{3} \cdot c_{p,w} \cdot \frac{T_{TES}(t+1,3) - T_{TES}(t,3)}{\Delta t} =$$
$$\dot{m}_{FCU} \cdot c_{p,w} \cdot T_{out}^{FCU}(t) - \dot{m}_{TES} \cdot c_{p,w} \cdot T_{TES}(t,3)$$
$$+(\dot{m}_{TES} - \dot{m}_{FCU}) \cdot c_{p,w} \cdot T_{TES}(t,2)$$
$$-k_{TES} \cdot \frac{A_{TES}}{3} \cdot [T_{TES}(t,3) - T_{amb}(t)]$$
$$+k_w \cdot A_{CS} \cdot \left[\frac{T_{TES}(t,2) - T_{TES}(t,3)}{h/3}\right] \quad (31)$$

where $k_{TES}$ is the thermal loss factor of the TESS, $k_w$ the thermal conductivity of water, $A_{CS}$ the cross section area of the TESS, $A_{TES}$ the surface area of the TESS, $m_{TES}$ the total mass of water in the TESS, and $h$ the height of the TESS.

The set of algebraic expressions (29)-(31) mathematically characterizes the evolution of the all segments of the TESS's temperature over time.

The temperature of the topmost segment of the TES system, i.e. $T_{TES}(t,1)$, is restricted to be lower than the maximum achievable temperature of the HP ($T_{in}^{TES,max}$). Therefore, the temperature limit of the topmost segment can be mathematically expressed as:

$$T_{TES}(t,1) \leq T_{in}^{TES,max} \quad (32)$$

Moreover, the temperature of each segment of the TESS is restricted to be lower than the temperature of its upper segment. Therefore, the temperature limit of segments 2 and 3 can be mathematically expressed as:

$$T_{TES}(t,i) \leq T_{TES}(t,i-1) \qquad i = 2,3 \quad (33)$$

In addition, the minimum temperature of TESS (i.e. the temperature of segment 3) must be greater than the flow temperature ($T_{Flow}(t)$) resulting from the building's heating curve to ensure thermal comfort.

$$T_{Flow}(t) \leq T_{TES}(t,3) \quad (34)$$

It should be noted that the inlet temperature of fan coil unit ($T_{in}^{FCU}(t)$) is equal to the temperature of the topmost segment ($T_{TES}(t,1)$) and the temperature of outlet water stream (from the bottom of TESS) to the HP's condenser is equal to the temperature of the lowest segment ($T_{TES}(t,3)$):

$$T_{in}^{FCU}(t) = T_{TES}(t,1) \quad (35)$$
$$T_{out}^{TES}(t) = T_{TES}(t,3) \quad (36)$$

Finally, the delivered thermal power from the TESS to the fan coil unit can be determined as:

$$P^{th,FCU}(t) = u^{FCU}(t) \cdot \dot{m}_{FCU} \cdot c_{p,w} \cdot [T_{in}^{FCU}(t) - T_{out}^{FCU}(t)] \quad (37)$$

where $u^{FCU}(t)$ is a binary variable representing whether the fan coil is working in time $t$ or not.

### E. The Heat Pump model

Heat pump (HP) is a key element of the building's energy system. This element is used to provide heating and cooling in the building. HP consumes electric power to transfer thermal energy from a source with a lower temperature to a sink with a higher temperature, accordingly HP can be operated in cooling or heating modes based on the required thermal load of the building. To enable the HP to maintain efficient operation under diverse environment conditions, a PVT-air dual source heat pump is used as shown in Fig. 8. This HP consists of a PVT evaporator and an air source evaporator connected in parallel that operate simultaneously to recover thermal energy from both solar energy and ambient air. In cases where the evaporating capacity in PVT evaporator remains relatively low under low solar irradiation, the air source evaporator can make amends for recovering heat. In literatures, there was rarely reported about dual source HP with double evaporators that encompasses air and solar source evaporators simultaneously [40,41]. The HP is modeled relying on its specifications presented in the datasheet. The thermal power absorbed in the PVT evaporator ($P^{th,evp,PVT}$) can be written as:

$$P^{th,EVP,PVT} = \dot{m}_{PVT}.c_{p,w}.(T_{out}^{PVT} - T_{evp}).$$
$$[1 - \exp\left(\frac{-UA_{evp,pvt}}{\dot{m}_{PVT}.c_{p,w}}\right)] \tag{38}$$

where, $UA_{evp,pvt}$ is the overall heat transfer coefficient of the PVT evaporator. The thermal power absorbed in the air source evaporator can be calculated as:

$$P^{th,EVP,air} = \dot{m}_{air}.c_{p,w}.(T_{amb} - T_{evp}).$$
$$[1 - \exp\left(\frac{-UA_{evp,air}}{\dot{m}_{air}.c_{p,w}}\right)] \tag{39}$$

where $UA_{evp,air}$ is the overall heat transfer coefficient of the air source evaporator. Therefore, the total thermal power absorbed in both evaporators is calculated as:

$$P^{th,EVP} = P^{th,EVP,PVT} + P^{th,EVP,air} \tag{40}$$

The HP coefficient of performance (COP) is expressed as:

$$COP = \frac{P^{th,HP}}{P^{e,HP}} \tag{41}$$

where $P^{th,HP}$ indicates the thermal power that HP delivers to the TESS in the condenser and $P^{e,HP}$ is the electric power needed to transfer thermal power from evaporators (with low temperature) to condenser (with higher temperature). Relying on expression (41), $P^{th,HP}$ can be expressed on the basis of $P^{e,HP}$:

$$P^{th,HP} = u^{HP}(t) \times COP \times P^{e,HP} \tag{42}$$

where $u^{HP}(t)$ is a binary variable representing whether the HP is working (i.e. is on) in time $t$ or not. The relation between HP electric power consumption, HP absorbed thermal power in evaporators and HP delivered thermal power to the TESS is:

$$P^{th,HP} = P^{e,HP} + P^{th,EVP} \tag{43}$$

Finally, the outlet temperature of water stream in load-side (i.e. condenser) of the HP can be calculated as:

$$T_{in}^{TES} = T_{out}^{TES} + \frac{P^{th,HP}}{\dot{m}_{TES}.c_{p,w}} \tag{44}$$

## III. Stochastic Model Predictive Control (MPC) Strategy

### A. Stochastic MPC framework and structure

This section constructs a stochastic MPC strategy to optimally steer the building's energy system over each individual day. More specifically, the MPC strategy starts at the beginning of the day and lasts until the end of the day, as illustrated in Fig. 9. Accordingly, the time horizon of the MPC problem (H) is 24 hours of the day which embraces two time periods:

- **Time slot** (T): The whole 24-hour of the day is split into 48 time slots with duration T = 30 minutes and $k$ =1, 2, ..., 48 indicates the rolling index for time slots over the day of operation. The MPC problem is solved at the beginning of each time slot, whereby, the operating set-point of the building's energy system is updated and actuated.

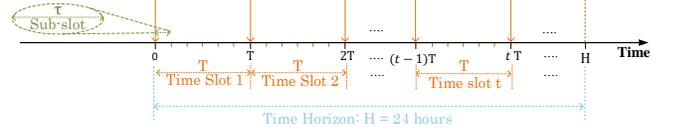

Fig. 9. Timeline of the MPC strategy over a day.

- **Sub-slot** ($\tau$): Each time slot consists of a number of sub-slots with duration of $\tau = 1$ minute and $\kappa = 1, 2, ..., 1440$ indicates the rolling index for sub-slots over the day of operation. The MPC problem models the trajectory of solar irradiation ($G$) with time resolution of $\tau = 1$ minutes to better capture the high frequency temporal variations of the solar irradiation. However, the outdoor ambient air temperature ($T_{amb}$) and electricity price ($\pi^{e,Grid}$) have moderate temporal variations, thus, their temporal variations are modeled with time resolution of T = 30 minutes.

The outline of the MPC strategy can be elaborated considering the situations at the beginning of time slot $k$:

1) The most recent realized outdoor ambient air temperature ($T_{amb}[k-1]$), building's indoor air temperature ($T_{in}[k-1]$) and solar irradiation ($G[\kappa-1]$) are measured.

2) the outdoor ambient air temperature ($T_{amb}[k]$, $T_{amb}[k+1]$, ..., $T_{amb}[48]$) and electricity price ($\pi^{e,Grid}[k]$, $\pi^{e,Grid}[k+1]$, ..., $\pi^{e,Grid}[48]$) over upcoming time slots ($k$, $k$ +1,..., 48) are predicated with time resolution of T = 30 minutes.

3) the solar irradiation ($G[\kappa]$, $G[\kappa+1]$, ..., $G[1440]$) over upcoming sub-slots ($\kappa$, $\kappa + 1$, ..., 1440) is predicted with time resolution of $\tau = 1$ minute[1]. Moreover, the prediction error of the solar irradiation ($\Delta G[\kappa,s]$, $\Delta G[\kappa + 1,s]$, ..., $\Delta G[1440,s]$) is modeled through $N_s$ scenarios with time resolution of $\tau = 1$ minute. It should be noted that $s$ =1, 2, ..., $N_s$ indicates the index of scenarios modeling predication error of solar irradiation

4) Value of all parameters and variables over each sub-slot can be retrieved from its value over the associated time slot (for all $\kappa = 30k, 30k + 1, ..., 30k + 29$):

$$\pi^{e,Grid}[\kappa] = \pi^{e,Grid}[k] \tag{45}$$
$$T_{amb}[\kappa] = T_{amb}[k] \tag{46}$$

5) The MPC strategy is actuated with time resolution of T = 30 minutes. To put it simply, at the beginning of time slot $k$, the MPC problem is solved considering all upcoming time slots ($k$, $k$ +1,..., 48) and accordingly the operating set-point of the HP and fan coils are determined over all the upcoming time slots. Then, the operating set-point of the HP and fan coils are updated/actuated for the first upcoming time slot. It should be noted that the building's energy system (i.e. HP and fan coils) can only take a unique operating set-point over each time slot (T = 30 minutes).

---

[1] Considering the timeline of the problem depicted in Fig. 9, $\kappa = 30k$.

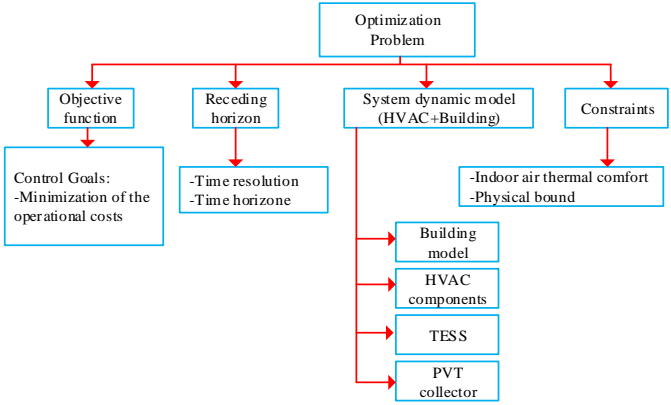

Fig. 10. General framework of the MPC optimization problem applied to a building and HVAC system.

### B. Formulation of the Stochastic MPC Strategy

The goal of the MPC strategy is to determine the optimal set-points of the HP and fan coil unit while minimizing the operation cost of the building's energy system over the whole upcoming time slots. The MPC strategy takes into account:

- the technical constraints of all elements of the building's energy system including PVT collector, HP and TESS.
- thermal comfort of the residents.
- temporal variations of the electricity price ($\pi^{e,Grid}$) and outdoor ambient air temperature ($T_{amb}$) over all upcoming time slots.
- Temporal variation of the solar irradiation ($G$) and its accompanied forecast error ($\Delta G$) over all upcoming time slots.

As shown in Fig. 10, the MPC problem is a holistic control approach comprising of different building blocks namely: the system dynamic model, objective function and constraints with a predefined time resolution and time horizon.

To mathematically formulate the MPC strategy, let's assume to be at the beginning of the time slot $k$.

#### 1) Objective of the MPC Strategy

The control objective is to minimize the operation cost of the building's energy system over the whole upcoming time slots ($k$, $k+1$,..., 48) and accordingly over the whole upcoming sub-slots ($\kappa$, $\kappa+1$, ..., 1440). This control objective can be formulated as:

$$min \sum_{\kappa \in \mathbb{K}} \sum_{s \in \mathbb{S}} \pi^{e,Grid}[\kappa]. P^{e,Grid}[\kappa, s] \qquad (47)$$

where $\mathbb{K} = \{\kappa, \kappa+1,..., 1440\}$ is the set of upcoming sub-slots, $\mathbb{S} = \{1, 2,..., N_s\}$ is the set of scenarios modeling the forecast error of solar irradiation, $P^{e,Grid}[\kappa, s]$ is the amount of electricity absorbed from the grid over the $\kappa^{th}$ sub-slot and scenario s. $\pi^{e,Grid}[\kappa]$ is the price of electricity over the $\kappa^{th}$ sub-slot. To put it another way, the MPC controller tries to switch the HP and fan coil on or off to shift the electric power consumption from peak to off-peak time slots, thereby, minimizing the purchased electricity cost form the grid.

#### 2) Constraints of the MPC strategy

The technical constraints of the TESS can be expressed as

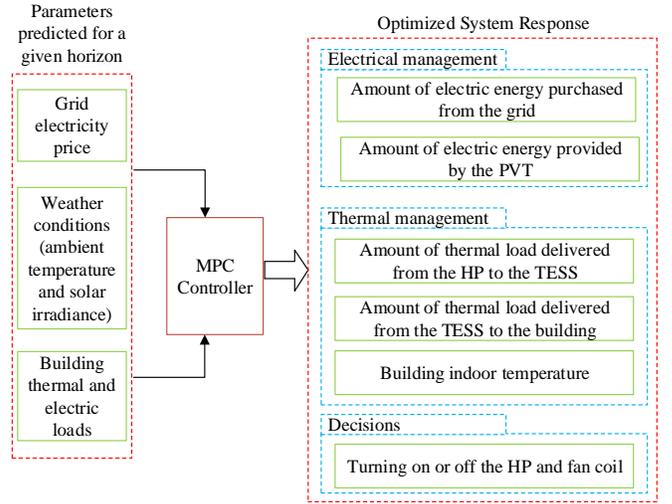

Fig. 11. The inputs and outputs of the MPC controller.

(for all $\kappa \in \mathbb{K} = \kappa, \kappa+1,..., 1440$ and $s \in \mathbb{S} = \{1, 2,..., N_s\}$):

$$T_{TES}[\kappa, s, j+1] \leq T_{TES}[\kappa, s, j] \qquad (48)$$
$$45^oC \leq T_{TES}[\kappa, s, j] \leq 95^oC \qquad (49)$$

where $\kappa$ is the index for the sub-slots and $s$ the index for scenarios modeling prediction error of solar irradiation, $j$ index for the segments of the TESS.

The electrical energy balance at the building can be expressed as (for all $\kappa \in \mathbb{K} = \{\kappa, \kappa+1,..., 1440\}$ and $s \in \mathbb{S} = \{1, 2,..., N_s\}$):

$$P^{e,building}[\kappa] + P^{e,HP}[\kappa, s] = P^{e,Grid}[\kappa, s] + P^{e,PVT}[\kappa, s] \quad (50)$$

The thermal energy balance at the building can be expressed as (for all $\kappa \in \mathbb{K} = \{\kappa, \kappa+1,..., 1440\}$ and $s \in \mathbb{S} = \{1, 2,..., N_s\}$):

$$P^{th,building}[\kappa, s] = P^{th,fancoil}[\kappa, s] \qquad (51)$$

To ensure the thermal comfort of the residents, the building indoor air temperature must be kept within an acceptable range (between 19ºC and 23ºC) over the whole upcoming sub-slots (for all $\kappa \in \mathbb{K} = \{\kappa, \kappa+1,..., 1440\}$ and $s \in \mathbb{S} = \{1, 2,..., N_s\}$):

$$19^oC \leq T_{in}[\kappa, s] \leq 23^oC \qquad (52)$$

Finally, the inputs and outputs of the MPC controller are visualized in Fig. 11. The MPC controller should predict and optimize energy utilization and operating costs, subject to changes in electricity prices, building thermal and electric loads, and weather conditions.

### C. Prediction model

Due to the volatility of the meteorological conditions, the predicted solar irradiation might differ from the one that realize during the real-time operation of the building. The difference between the predicted value and the realized value is called prediction error or uncertainty. Considering the fact that the solar irradiation features high frequency variations, the method models the solar irradiation through a set of scenarios, i.e. $\mathbb{S}$, to account for the prediction errors. Moreover, it models each scenario with time resolution of 1-minute to better capture the temporal variations of the solar irradiation over each time slot. This 1-minute duration is called sub-slot as shown in Fig. 9. Accordingly, the uncertainty-cognizant MPC strategy is able to better ensure the thermal comfort of the residents in comparison with common MPC strategies that generally model the solar irradiation with a single scenario (i.e. point prediction) and long-time resolutions range between 30-minute to 1-hour. To model the set of scenarios 𝕊, the paper relies on a *Training Data Set* consisting of a series of historical measured solar irradiation ($G_T$ (w/m²)) over the last 10 years with time resolution of 1 minute. Then, a machine learning based approach, i.e. k-nearest neighbor (k-NN) algorithm [42], is exploited to generate 100 scenarios modeling the solar irradiation over the upcoming sub-slots ($\kappa$, $\kappa$ + 1, …, 1440).

Table 1. Corresponding parameters of the building thermal model.

| Parameter | Value |
|---|---|
| $R_{wall,ex}$ ($m^2k/W$) | 0.0924 |
| $R_{wall}$ ($m^2k/W$) | 0.0853 |
| $R_{wall,in}$ ($m^2k/W$) | 0.004 |
| $R_{win}$ ($m^2k/W$) | 0.0102 |
| $R_i$ ($m^2k/W$) | 0.0065 |
| $C_{wall,ex}$ ($J/m^2K$) | $205*10^3$ |
| $C_{wall,in}$ ($J/m^2K$) | $195*10^3$ |
| $C_{in}$ ($J/m^2K$) | $3.69*10^3$ |
| $C_{itm}$ ($J/m^2K$) | $526*10^3$ |

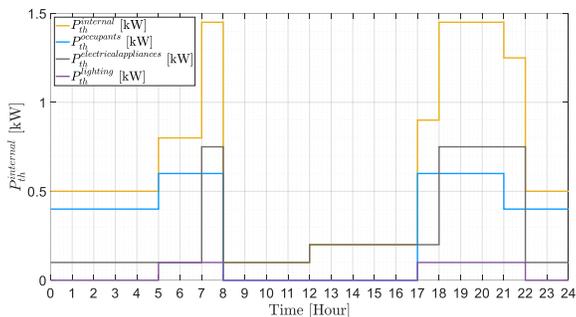

Fig. 12. The internal heat load of examined building.

Table 2. Electrical and thermal performance coefficients of the PVT collector.

| Parameter | Unit | Value |
|---|---|---|
| $A_{PVT}$ | $m^2$ | 8 |
| $\eta_o$ | - | 0.513 |
| $K_d$ | - | 0.903 |
| $b_{o,th}$ | - | 0.087 |
| $C_1$ | W/$m^2$K | 6.032 |
| $C_2$ | W/$m^2K^2$ | 0.035 |
| $C_3$ | J/$m^3$K | 0.00008 |
| $C_4$ | - | 0.203 |
| $C_5$ | J/$m^2$K | 16912 |
| $C_6$ | (m/s)$^{-1}$ | 0.006 |
| $R_{PVT}$ | $m^2$K/W | 0.0662 |
| $b_{0,el}$ | - | 0.238 |
| $\eta_{el,ref}$ | - | 0.1228 |
| $\beta$ | %/K | 0.370 |
| $a$ | $m^2$/W | 0.00001090 |
| $b$ | - | -0.04700 |
| $c$ | - | -1.400 |

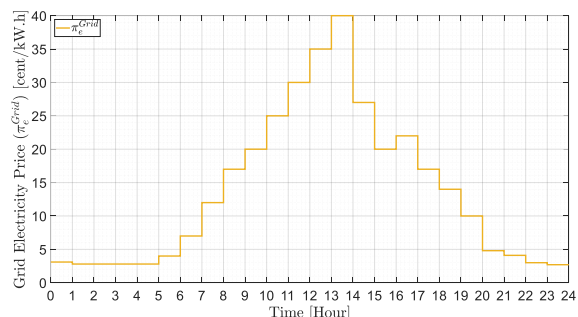

Fig. 13. Hourly real time pricing profile of grid electricity.

Considering the fact that the ambient temperature has a very moderate variation over time, this method only accounts for the point prediction of the ambient temperature over the upcoming time slots and it does not account for the prediction error of the ambient temperature. Moreover, this method models the ambient temperature with time resolution equal to the duration of a time slot, i.e. 30 minutes. In this respect, this paper relies on a *Training Data Set* consisting of a series of historical measured ambient temperature over the last 10 years with time resolution of 15 minutes. It is worth highlighting that this data set has been received from Iran Meteorological Organization (IMO). Then, k-NN algorithm is exploited to predict the ambient temperature over the upcoming time slots ($k$, $k$ +1,…, 48).

## IV. CASE STUDY

The performance of the proposed uncertainty-cognizant MPC is investigated considering a single-family residential building with 50 m² area over a winter day. The parameters of the building thermal model, i.e. resistances and capacitances introduced in section II.B, are reported in Table 1. It is worth highlighting that these parameters are calculated by applying a GA-based parameter identification approach on historical thermal data of a typical building in Tehran, Iran [43]. This study is also run considering the internal heat load of the building ($P_{th}^{internal}$) comprising of the occupant heat gain ($P_{th}^{occupants}$), electrical appliances heat gain ($P_{th}^{electrical\ appliances}$), and lightning heat gain ($P_{th}^{lighting}$). The profiles of all these terms over 24 hours of study are shown in Fig. 12.

The building's energy system is formed of 1-a PVT-air dual source heat pump with R-32 refrigerant, 2-a 2000-liter water storage tank with a height-over-diameter ratio of 4 as a TESS, 3-a water-based covered photovoltaic thermal (PVT) collector with front glazing, rear collector cover and no thermal insulation material on the back of the PVT absorber, 4-circulation pumps and 5-FCUs. The performance coefficients of the PVT collector are reported in Table 2 [31,32].

Considering the fact that the hourly electricity prices are determined in the day-ahead electricity market (a day prior to real-time operation), this study assumes that the hourly electricity prices are known/given. More specifically, the hourly electricity prices are extracted from the published data by Iran Electricity Market (IEMA). The profile of the hourly electricity prices over 24 hours of the study, i.e. 10 January 2021, is shown in Fig. 13.

All prediction and simulations are carried out using MATLAB. Moreover, the stochastic MPC optimization problem is modeled by using YALMIP-MATLAB and solved with GUROBI solver on a Windows based system with a AMD Ryzen 5 3500U CPU and 16 GB RAM.

## V. RESULTS AND DISCUSSION

### A. Simulation results associated with uncertainty-cognizant MPC strategy (Controller 1)

This section presents the performance of the constructed uncertainty-cognizant MPC strategy for managing the building energy system over a typical winter day. At the beginning of each time slot, the solar irradiation and ambient air temperature are firstly predicted following the method presented in section III.C. Fig. 14 depicts all 100 scenarios modeling the solar irradiation (over the upcoming sub-slots) along with the realized solar irradiation over the 24 hours of the study. Moreover, the point prediction of the solar irradiation is calculated so that fifteen percent of scenarios are above it and the 50 percent of scenarios are below it, i.e. quantile 50%. Fig. 15 illustrates the predicted ambient air temperature along with the realized one. Considering the predicted solar irradiation as well as ambient temperature along with the model developed in section II.C, the thermal and electric power generation of the PVT collector can be easily calculated (predicted) as shown in Fig. 16. Finally, the stochastic MPC strategy presented in section III are actuated at the beginning of each time slot to optimally steer the building's energy system. Accordingly, the optimal operating points of the HP and FCU are determined. Fig. 17 and Fig. 18 respectively show the HP and FCU operating points (i.e. off/on state) along with their output thermal power. Fig. 19 illustrates the evolution of the building's indoor air temperature and TESS water temperature over the whole day. As it can be seen, the method manages to keep the indoor air temperature and TESS water temperature within their desired ranges, thereby, preserving the thermal comfort of the residents. Moreover, from 00:00 to 06:00 where the electricity price is low, the method repeatedly turns on the HP and FCU to store thermal energy in the thermal capacities of the TESS and building (i.e. increases the TESS water temperature and keep the indoor air temperature higher than the minimum acceptable indoor air temperature). To better discover the capability of the method in optimally steering the building's energy system, Fig. 20 shows the net electric power that the building exchanged with the grid over 24 hours of study. In Fig. 20, positive values mean that the building absorbed (i.e. purchased) electricity from the grid, while the negative values mean that the building injected (i.e. sold) electricity to the grid. As it can be seen, the method leverages the thermal capacities of the TESS and building to shift the building's electricity consumption from the periods with high prices to periods with low prices, thereby, decreasing the operating cost of the building's energy system. More specifically, the method turns on the HP during periods with low prices, thereby, storing thermal energy in the TESS. Then, during periods with high prices (9:00 am and 3:00 pm), the method turns off the HP and discharges the TESS. In this way, the building is empowered to sell electricity to the grid during periods with high prices and consume electricity mainly during periods with low prices.

The building's net purchased electricity over the whole day is 164 kWh (i.e. equal to the area under the curve shown in Fig. 20). Considering the variable electricity price over the day, the building's electricity cost over the whole day is 221 cents.

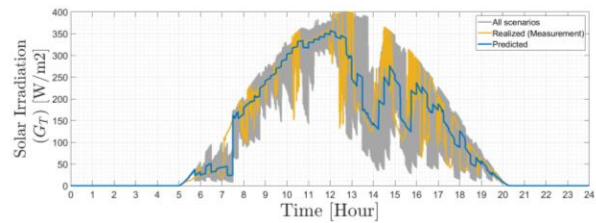

Fig. 14. The realized and predicted solar irradiation on the examined day.

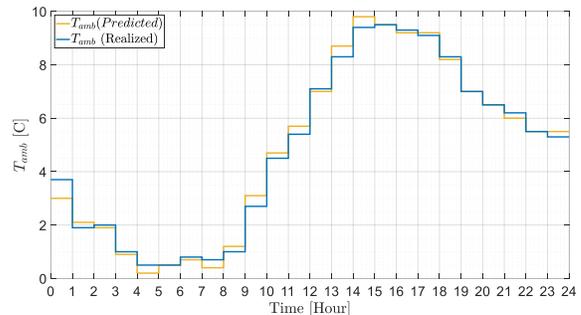

Fig. 15. The realized and predicted ambient air temperature on the examined day.

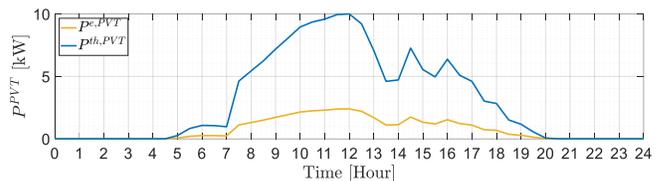

Fig. 16. The electric and thermal power generated by the PVT collector.

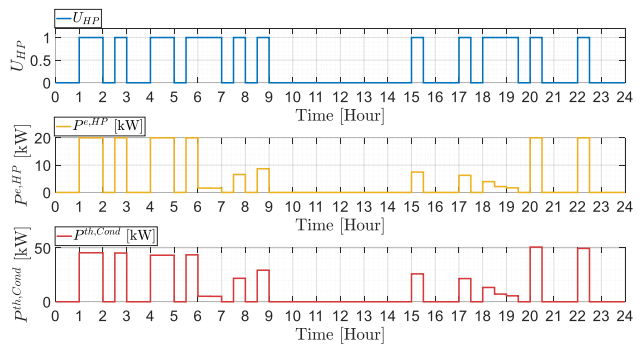

Fig. 17. Controller 1: The HP off/on state (0 or 1 binary variables), HP consumed electric power ($P^{e,HP}$), and HP output thermal power ($P^{th,HP}$) over 24 hours of study.

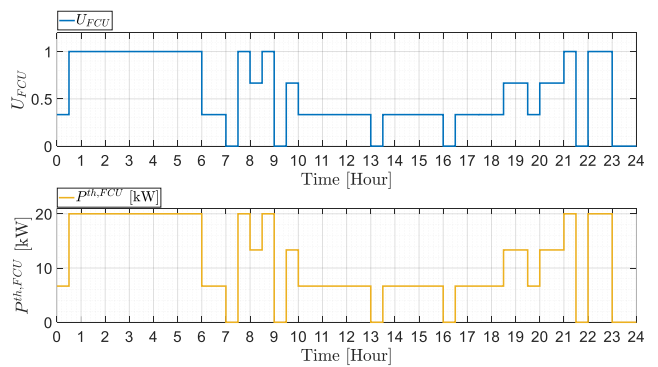

Fig. 18. Controller 1: The FCU off/on state (0 or 1 binary variables) and FCU thermal power ($P^{th,FCU}$) over 24 hours of study.

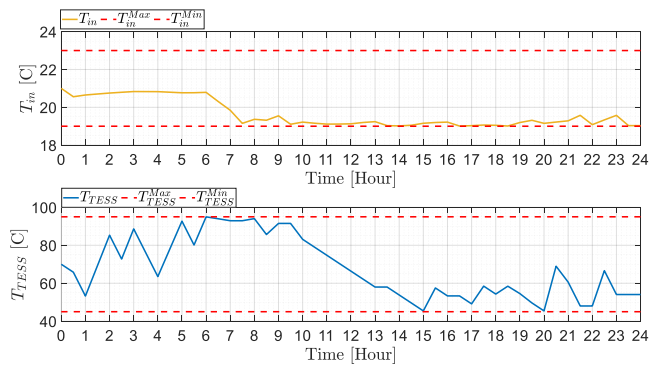

Fig. 19. Controller 1: Evolution of the building's indoor air ($T_{in}$) temperature and TESS water temperature ($T_{TES}$) over 24 hours of study.

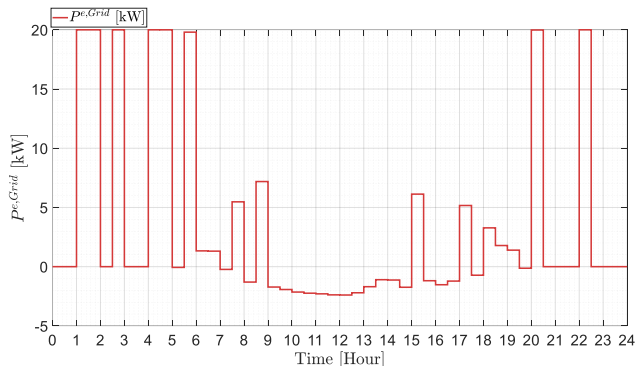

Fig. 20. Controller 1: The net electric power sold to (negative value) or purchased from (positive value) from the grid over 24 hours of study.

### B. Simulation results associated with thermostatic controller (Controller 2)

This section aims to better unfold the advantages of the constructed uncertainty-cognizant MPC strategy (Controller 1) by comparing its performance with the thermostatic controllers (Controller 2) that are widely used to manage the building's energy system. To this end, it is assumed that two thermostatic controllers (Controller 2) are used to manage the building's energy system, i.e. FCU and HP. These thermostatic controllers (Controller 2) can allow the residents to only determine a single set-point for the indoor air temperature, i.e. $T_{in}^{Set-point}$, and TESS water temperature, i.e. $T_{TESS}^{Set-point}$, instead of allowing the residents to determine a range for the indoor air temperature and TESS water temperature. A hysteresis controller is used to control the FCU. This controller turns on the FCU when the indoor air temperature goes bellow the determined setpoint $T_{in}^{Set-point}$ and vice versa. The other controller is in charge of the HP. This controller turns on the HP when the TESS water temperature goes bellow the determined setpoint $T_{TESS}^{Set-point}$ and vice versa. The performance of the thermostatic controller (i.e. Controller 2) is completely dependent on the determined set-points $T_{in}^{Set-point}$ and $T_{TESS}^{Set-point}$. This section sets these set-points (i.e. $T_{in}^{Set-point}$ and $T_{TESS}^{Set-point}$) via try and error so that the indoor air temperature is kept as low as possible while:

1-the thermal comfort of the residents is respected throughout the 24 hours of the study, i.e. $T_{in}$ stays between 19ºC and 23ºC.

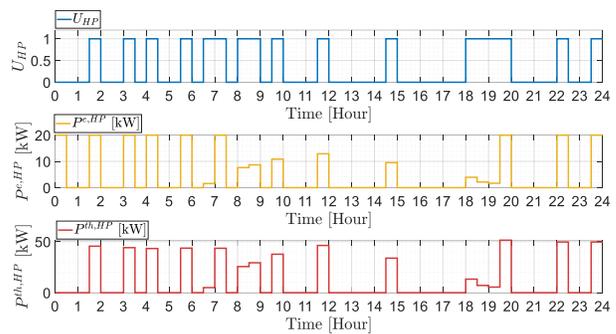

Fig. 21. Controller 2: The HP off/on state (0 or 1 binary variables), HP consumed electric power ($P^{e,HP}$), and HP output thermal power ($P^{th,HP}$) over 24 hours of study.

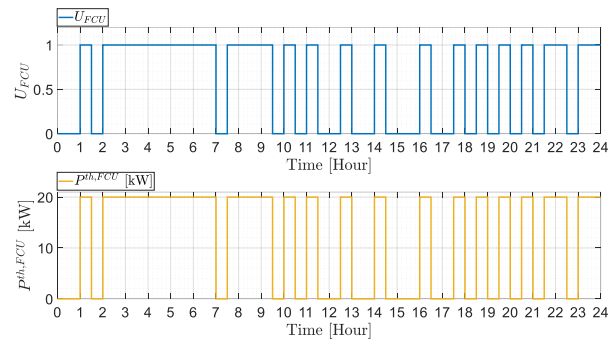

Fig. 22. Controller 2: The FCU off/on state (0 or 1 binary variables) and FCU thermal power ($P^{th,FCU}$) over 24 hours of study.

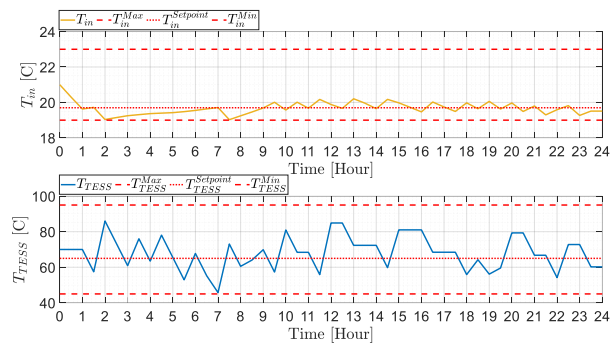

Fig. 23. Controller 2: Evolution of the building's indoor air temperature ($T_{in}$) and TESS water temperature ($T_{TES}$) over 24 hours of study.

2-$T_{TES}$ stays between $T_{TESS}^{Min}$ (45ºC) and $T_{TESS}^{Max}$ (95ºC) respected throughout the 24 hours of the study.

In this way, the thermostatic controllers are able to steer the building's energy system with minimum cost while preserving the thermal comfort of the residents. Therefore, a fair comparison can be drawn between Controller 1 and Controller 2. Following the above-mentioned try and error procedure, $T_{in}^{Set-point}$ and $T_{TESS}^{Set-point}$ are respectively set 19.7 ºC and 65 ºC. The performance of the thermostatic controllers (i.e. Controller 2) are shown in Fig. 21 to Fig. 24. Fig. 21 and Fig. 22 respectively show the HP and FCU operating points (i.e. off/on state) along with their output thermal power. Fig. 23 illustrates the evolution of the building's indoor air temperature and TESS water temperature over the whole day. As it can be seen, the thermostatic controller keeps the building's indoor air temperature around the set-point 19.7 ºC. Fig. 24 shows the net electric power that the building exchanged with the grid over 24 hours of study. In this figure, positive values mean that the

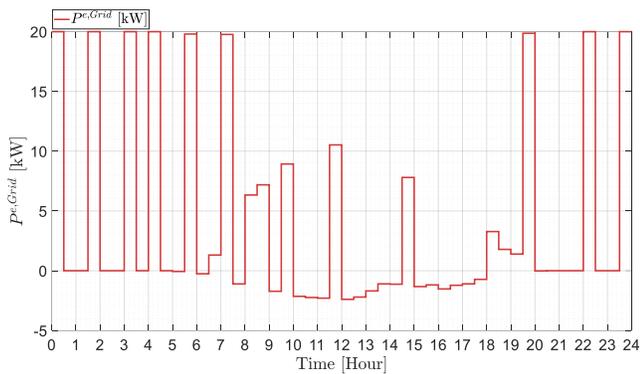

Fig. 24. Controller 2: The net electric power sold to (negative value) or purchased from (positive value) from the grid over 24 hours of study.

Table 3. Comparison of total electric power consumption and electricity cost for the two considered cases.

| Controller | Total energy consumption | | Electricity cost | |
|---|---|---|---|---|
| | kW.h | %* | cent | %* |
| Controller 1: uncertainty-cognizant MPC | 164 | 31% | 221 | 81% |
| Controller 2: thermostatic controller | 239 | - | 1215 | - |

* Percentage of improvement in respect to controller 2.

building absorbed (i.e. purchased) electricity from the grid, while the negative values mean that the building injected (i.e. sold) electricity to the grid. As it can be seen, the thermostatic controller (i.e. Controller 2) never can shift the building's electricity consumption from the periods with high prices to periods with low prices because:

1- the thermostatic controller solely relies on the real-time measurements,

2- the thermostatic controller is not able to take advantage from the prediction of solar irradiation and ambient air temperature, thereby, cannot take advantage from the thermal capacities of the TESS.

To put it simply, the thermostatic controller (i.e. Controller 2) cannot decrease the operating cost of the building's energy system.

Relying on the thermostatic controller (i.e. Controller 2), the building's net purchased electricity over the whole day is 239 kWh (i.e. equal to the area under the curve shown in Fig. 24). Considering the variable electricity price over the day, the building's electricity cost over the whole day is 1,215 cents.

To better unfold the advantage of controller 1, Table 3 summaries the performance of the uncertainty-cognizant MPC strategy (i.e. Controller 1) and thermostatic controller (i.e. Controller 2). This table reveals that controller 1 results in 31% less electric power consumption and 81% electricity cost saving in comparison with controller 2. It is worth highlighting that controller 1 achieved a much better performance due to 1-taking advantage from the thermal capacities of the TESS and the building 2-accounting for the temporal variation of the electricity price, solar irradiation and ambient temperature over the upcoming time-slots.

## VI. CONCLUSIONS

This paper advocates for the employment of uncertainty-cognizant model predictive control (MPC) strategy for optimally operating smart buildings equipped with grid-connected photovoltaic-thermal (PVT) collectors, heat pump (HP) system, thermal energy storage system (TESS) and fan coil unit (FCU). In this context, it constructed a stochastic MPC strategy to maximize the building's self-consumption and consequently minimize building's electricity cost while considering the temporal variations of the electricity price (i.e. time-of-use (TOU) tariff), renewable energy resources (RES) generation, and outdoor ambient air temperature over the whole day. More specifically, the presented method minimizes the building's electric power consumption (and more importantly the building's electricity cost) while preserving the thermal comfort of the residents. The uncertainties associated with RES were also modeled by k-NN algorithm through defining a set of scenarios to strengthen the modeling the RES power generation. A single-family residential building is used to validate the performance of the proposed method. The achieved results show that the proposed stochastic MPC significantly decreases the building's total electricity consumption about 75 kWh (31%) resulting in 991 cent (81%) energy cost saving by shifting the HP and FCU loads while keeping the indoor temperature within its allowable temperature range. It is of great importance to note that the proposed method takes the advantage of the thermal energy storage capacity to shift the building's energy consumption from peak to off-peak hours, thereby, reducing the building's electricity cost over the whole day.

REFERENCES

[1] Omer, A.M., 2008. Energy, environment and sustainable development. Renewable and sustainable energy reviews, 12(9), pp.2265-2300.

[2] Gielen, D., Boshell, F., Saygin, D., Bazilian, M.D., Wagner, N. and Gorini, R., 2019. The role of renewable energy in the global energy transformation. Energy Strategy Reviews, 24, pp.38-50.

[3] Röck, M., Saade, M.R.M., Balouktsi, M., Rasmussen, F.N., Birgisdottir, H., Frischknecht, R., Habert, G., Lützkendorf, T. and Passer, A., 2020. Embodied GHG emissions of buildings–The hidden challenge for effective climate change mitigation. Applied Energy, 258, p.114107.

[4] Tan, D. and Novosel, D., 2017. Energy challenge, power electronics & systems (PEAS) technology and grid modernization. CPSS Transactions on Power Electronics and Applications, 2(1), pp.3-11.

[5] Hedman, Å., Rehman, H.U., Gabaldón, A., Bisello, A., Albert-Seifried, V., Zhang, X., Guarino, F., Grynning, S., Eicker, U., Neumann, H.M. and Tuominen, P., 2021. IEA EBC Annex83 positive energy districts. Buildings, 11(3), p.130.

[6] International Energy Agency (IEA) Renewables-Fuels & Technologies. Available online: https://www.iea.org/fuels-andtechnologies/renewables (accessed on 6 October 2020).

[7] Pau, G., Collotta, M., Ruano, A. and Qin, J., 2017. Smart home energy management.

[8] Martin-Chivelet, N. and Montero-Gomez, D., 2017. Optimizing photovoltaic self-consumption in office buildings. Energy and Buildings, 150, pp.71-80.

[9] Luthander, R., Widén, J., Nilsson, D. and Palm, J., 2015. Photovoltaic self-consumption in buildings: A review. Applied energy, 142, pp.80-94.

[10] Arteconi, A., Hewitt, N.J. and Polonara, F., 2013. Domestic demand-side management (DSM): Role of heat pumps and thermal energy storage (TES) systems. Applied thermal engineering, 51(1-2), pp.155-165.

[11] Kim, N.K., Shim, M.H. and Won, D., 2018. Building energy management strategy using an HVAC system and energy storage system. Energies, 11(10), p.2690.

[12] American Society of Heating, Refrigerating and Air-Conditioning Engineers (ASHRAE). 2013 ASHRAE Handbook—Fundamentals (SI Edition); American Society of Heating, Refrigerating and Air-Conditioning Engineers, Inc.: Atlanta, GA, USA, 2013

[13] American Society of Heating, Refrigerating and Air-Conditioning Engineers (ASHRAE). 2015 ASHRAE Handbook—HVAC Applications; American Society of Heating, Refrigerating and Air-Conditioning Engineers, Inc.: Atlanta, GA, USA, 2015.

[14] Serale, G., Fiorentini, M., Capozzoli, A., Bernardini, D. and Bemporad, A., 2018. Model predictive control (MPC) for enhancing building and HVAC system energy efficiency: Problem formulation, applications and opportunities. Energies, 11(3), p.631.

[15] Afram, A. and Janabi-Sharifi, F., 2014. Theory and applications of HVAC control systems–A review of model predictive control (MPC). Building and Environment, 72, pp.343-355.

[16] Tang, R. and Wang, S., 2019. Model predictive control for thermal energy storage and thermal comfort optimization of building demand response in smart grids. Applied Energy, 242, pp.873-882.

[17] Halvgaard, R., Bacher, P., Perers, B., Andersen, E., Furbo, S., Jørgensen, J.B., Poulsen, N.K. and Madsen, H., 2012. Model predictive control for a smart solar tank based on weather and consumption forecasts. Energy Procedia, 30, pp.270-278.

[18] Deng, K., Sun, Y., Li, S., Lu, Y., Brouwer, J., Mehta, P.G., Zhou, M. and Chakraborty, A., 2014. Model predictive control of central chiller plant with thermal energy storage via dynamic programming and mixed-integer linear programming. IEEE Transactions on Automation Science and Engineering, 12(2), pp.565-579.

[19] D'Ettorre, F., Conti, P., Schito, E. and Testi, D., 2019. Model predictive control of a hybrid heat pump system and impact of the prediction horizon on cost-saving potential and optimal storage capacity. Applied Thermal Engineering, 148, pp.524-535.

[20] Carvalho, A.D., Moura, P., Vaz, G.C. and de Almeida, A.T., 2015. Ground source heat pumps as high efficient solutions for building space conditioning and for integration in smart grids. Energy conversion and management, 103, pp.991-1007.

[21] Baniasadi, A., Habibi, D., Bass, O. and Masoum, M.A., 2018. Optimal real-time residential thermal energy management for peak-load shifting with experimental verification. IEEE Transactions on Smart Grid, 10(5), pp.5587-5599.

[22] Touretzky, C.R. and Baldea, M., 2014. Integrating scheduling and control for economic MPC of buildings with energy storage. Journal of Process Control, 24(8), pp.1292-1300.

[23] Shah, J.J., Nielsen, M.C., Shaffer, T.S. and Fittro, R.L., 2015. Cost-optimal consumption-aware electric water heating via thermal storage under time-of-use pricing. IEEE Transactions on Smart Grid, 7(2), pp.592-599.

[24] Alimohammadisagvand, B., Jokisalo, J., Kilpeläinen, S., Ali, M. and Sirén, K., 2016. Cost-optimal thermal energy storage system for a residential building with heat pump heating and demand response control. Applied Energy, 174, pp.275-287.

[25] Salakij, S., Yu, N., Paolucci, S. and Antsaklis, P., 2016. Model-Based Predictive Control for building energy management. I: Energy modeling and optimal control. Energy and Buildings, 133, pp.345-358.

[26] Avci, M., Erkoc, M., Rahmani, A. and Asfour, S., 2013. Model predictive HVAC load control in buildings using real-time electricity pricing. Energy and Buildings, 60, pp.199-209.

[27] Andersen, K.K., Madsen, H. and Hansen, L.H., 2000. Modelling the heat dynamics of a building using stochastic differential equations. Energy and Buildings, 31(1), pp.13-24.

[28] Madsen, H., 2007. Time series analysis. Chapman and Hall/CRC.

[29] Madsen, H. and Holst, J., 1995. Estimation of continuous-time models for the heat dynamics of a building. Energy and buildings, 22(1), pp.67-79.

[30] Vrettos, E., Kara, E. C., MacDonald, J., Andersson, G., & Callaway, D. S. (2016). Experimental demonstration of frequency regulation by commercial buildings—Part I: Modeling and hierarchical control design. IEEE Transactions on Smart Grid, 9(4), 3213-3223.

[31] Jonas, D., Lämmle, M., Theis, D., Schneider, S. and Frey, G., 2019. Performance modeling of PVT collectors: Implementation, validation and parameter identification approach using TRNSYS. Solar Energy, 193, pp.51-64.

[32] Lämmle, M., Oliva, A., Hermann, M., Kramer, K. and Kramer, W., 2017. PVT collector technologies in solar thermal systems: A systematic assessment of electrical and thermal yields with the novel characteristic temperature approach. Solar Energy, 155, pp.867-879.

[33] Duffie, J.A. and Beckman, W.A., 2013. Solar Engineering of Thermal Processes. John Wiley & Sons.

[34] Heydenreich, W., Müller, B. and Reise, C., 2008, September. Describing the world with three parameters: a new approach to PV module power modelling. In 23rd European PV Solar Energy Conference and Exhibition (EU PVSEC) (pp. 2786-2789).

[35] Skoplaki, E. and Palyvos, J.A., 2009. On the temperature dependence of photovoltaic module electrical performance: A review of efficiency/power correlations. Solar energy, 83(5), pp.614-624.

[36] Schütz, T., Streblow, R. and Müller, D., 2015. A comparison of thermal energy storage models for building

energy system optimization. Energy and Buildings, 93, pp.23-31.

[37] Han, Y.M., Wang, R.Z. and Dai, Y.J., 2009. Thermal stratification within the water tank. Renewable and Sustainable Energy Reviews, 13(5), pp.1014-1026.

[38] Li, Z.F. and Sumathy, K., 2002. Performance study of a partitioned thermally stratified storage tank in a solar powered absorption air conditioning system. Applied Thermal Engineering, 22(11), pp.1207-1216.

[39] Karim, M.A., 2011. Experimental investigation of a stratified chilled-water thermal storage system. Applied Thermal Engineering, 31(11-12), pp.1853-1860.

[40] Shan, M., Yu, T.H. and Yang, X., 2016. Assessment of an integrated active solar and air-source heat pump water heating system operated within a passive house in a cold climate zone. Renewable energy, 87, pp.1059-1066.

[41] Cai, J., Ji, J., Wang, Y. and Huang, W., 2017. Operation characteristics of a novel dual source multi-functional heat pump system under various working modes. Applied Energy, 194, pp.236-246.

[42] Chu, Y. and Coimbra, C.F., 2017. Short-term probabilistic forecasts for direct normal irradiance. Renewable Energy, 101, pp.526-536.

[43] Wang, S., & Xu, X., 2006. Simplified building model for transient thermal performance estimation using GA-based parameter identification. International journal of thermal sciences, 45(4), pp.419-432.